\documentclass{JAC2003}
\pdfoutput=1
\usepackage[pdftex]{graphicx}
\usepackage{booktabs}
\usepackage{stkernel}
\usepackage[T1]{fontenc}
\usepackage{mathptmx}
\usepackage[english]{babel}
\usepackage[spacing,tracking,kerning,letterspace=30,babel,stretch=50,shrink=50]{microtype}
\usepackage{textcomp}

\makeatletter
\renewcommand{\section}{\@startsection {section}{1}{0mm}{9pt}{4pt}{\centering\scshape\bfseries\fontsize{12}{16}\selectfont\MakeUppercase}}
\renewcommand{\subsection}{\@startsection {subsection}{2}{0mm}{6pt}{4pt}{\raggedright\itshape\fontsize{12}{16}\selectfont}}
\makeatother

\fussy

\begin{document}
\title{\textsc{GAS ELECTRON MULTIPLIERS\linebreak[1] FOR THE ANTIPROTON DECELERATOR}}
\author{S.\,Duarte Pinto\thanks{Serge.Duarte.Pinto@cern.ch}, R.\,Jones, L.\,Ropelewski, J.\,Spanggaard, G.\,Tranquille, \textsc{\textsc{cern}}, Geneva, Switzerland}

\maketitle

\begin{abstract}
The new beam profile measurement for the Antiproton Decelerator (\textsc{ad}) at \textsc{cern} is based on a single Gas Electron Multiplier (\textsc{gem}) with a 2D readout structure.
This detector is very light ($\sim0.4$\% X$_0$), and measures horizontal and vertical profiles directly in one plane.
This overcomes the problems previously encountered with multi-wire proportional chambers for the same purpose, where beam interactions with the detector severely affect the obtained profiles.
A prototype was installed and successfully tested in late 2010, with another five detectors now installed in the \textsc{asacusa} and \textsc{aegis} beam lines.
This paper will provide a detailed description of the detector and discuss the results obtained.
\end{abstract}

\section{Introduction}
\fnbelowfloat
The antiproton decelerator at \textsc{cern} delivers antiproton beams of two different energies to five experiments.
The antihydrogen experiments \textsc{alpha, asacusa, atrap} and \textsc{aegis} receive 5.3~MeV kinetic energy ($p=100$ MeV/c) beams, while the \textsc{ace} experiment for cancer therapy uses beams of 126 MeV (p=502 MeV/c) \cite{AD}.
The beam is extracted to one of the experiments in a spill of a few hundred na\-no\-se\-conds, containing about $3\cdot10^7$ antiprotons.
The \linebreak[4] leng\-thy deceleration sequence imposes a delay of almost two minutes between spills.

\begin{figure}[b]
\includegraphics[width=\columnwidth]{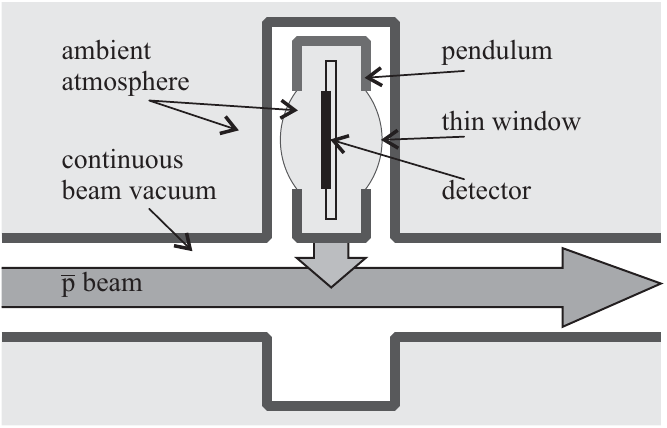}
\caption{Situation of a profile detector in a pendulum.}
\label{pendulum}
\end{figure}
Transverse profile information is needed at several locations along the extraction lines, to optimize the transfer optics.
It is very difficult to measure such profiles without ab\-sorbing a significant fraction of the beam.
Therefore, the detector is installed in a pendulum that can move through the beam vacuum (see Fig.~\ref{pendulum}).
The inside of the pendulum is in contact with ambient atmosphere; thin stainless steel windows separate it from the vacuum.
When the pendulum is positioned in the beam, the beam is entirely absorbed while measuring a profile.
When experiments are taking data, all pendulums are in the stalled position (as in Fig.~\ref{pendulum}) and the beam traverses an uninterrupted vacuum.
The tube that connects the inside of the pendulum with the outside atmosphere also serves as feed-trough for signal lines, high voltage cables, gas pipes, and a compressed air inlet that is used to cool the detector during a vacuum bake-out.

\section{The Detector}
\begin{figure*}
\includegraphics[width=\textwidth]{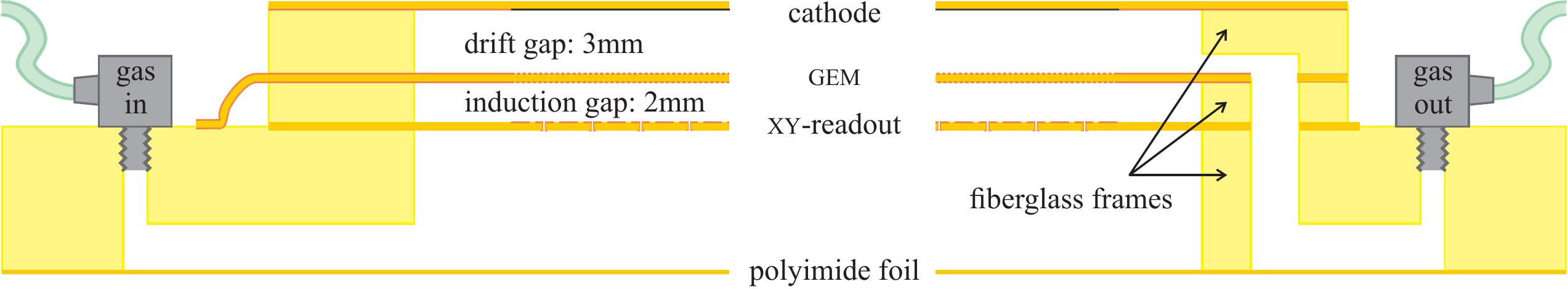}
\caption{Schematic buildup of the detector, showing components, gas features and dimensions.}
\label{detector}
\end{figure*}
The detectors installed in the pendulums are gaseous radiation detectors based on a single gas electron multiplier (\textsc{gem}).
The active area is $10\times10$ cm$^2$.
Figure~\ref{detector} shows schematically a cross-section of a detector; the \textsc{gem} foil is indicated between the cathode and the readout pattern.
The design of these chambers is optimized for low material budget, as the low energy beam is easily affected by multiple scattering.
Even materials downstream the active volume are kept light to minimize the effect of \emph{backsplash} from antiproton-nucleus annihilations downstream on the measured profiles.
The whole detector presents about 0.40\%~X$_0$ of material to the incoming beam.
The upstream vacuum window of the pendulum is located about 4 cm away from the detector, and adds 0.12\%~X$_0$.

\subsection{Gas Electron Multipliers}
\begin{figure}
\includegraphics[width=\columnwidth]{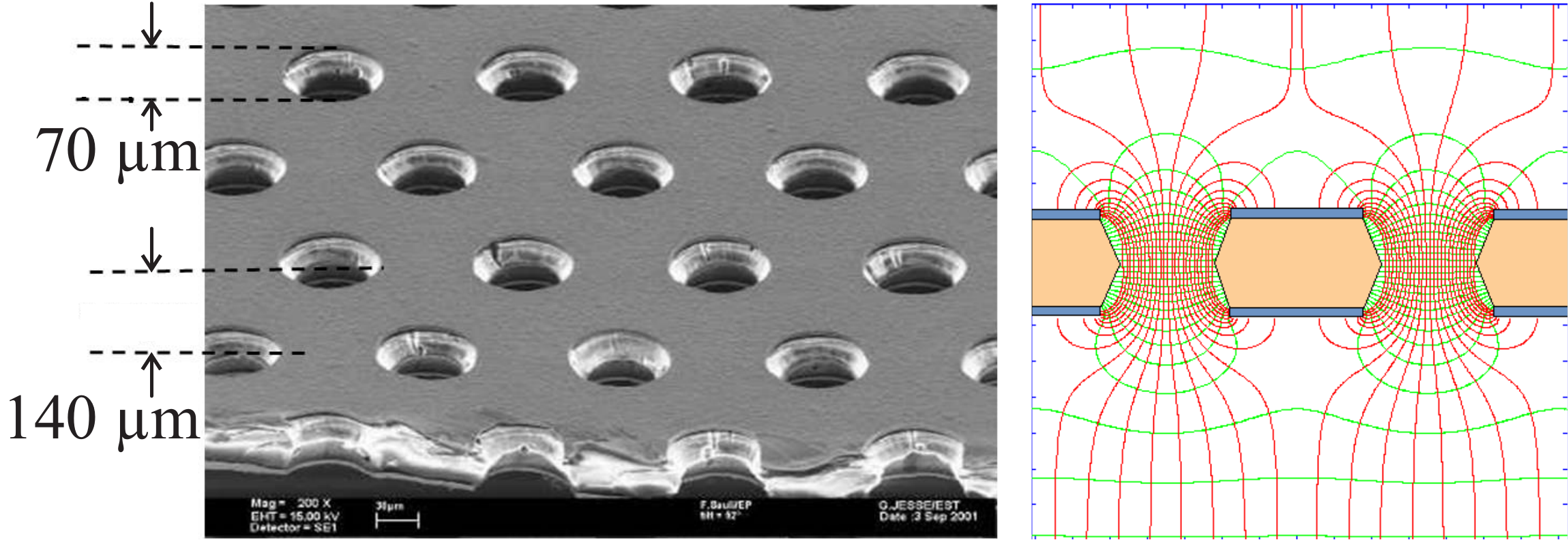}
\caption{Left: an electron micrograph of a \textsc{gem} foil, indicating hole dimensions and spacing. Right: calculated field pattern inside and around a \textsc{gem} hole.}
\label{GEM}
\end{figure}

A \textsc{gem} is a polyimide foil, copperclad on both sides, and pierced with a high density of holes \cite{firstGEM}.
The foil has a thickness of 50 \textmu m and holes are typically spaced with a pitch of 140 \textmu m in a triangular pattern, as indicated in Fig.~\ref{GEM} (left).
When a voltage is applied between top and bottom electrodes, an electric field is formed that strongly focuses in the center of the holes.
Inside the holes the field is orders of magnitude stronger than outside, this can be seen from the field map on the right of Fig.~\ref{GEM}.
In this strong field gas amplification takes place.

When ionization electrons from interactions of beam particles with gas molecules drift toward a \textsc{gem}, the charge is multiplied by a gain up to a few thousand per \textsc{gem}, and this charge is extracted from the other side of the \textsc{gem} to be collected by the readout circuit.
\textsc{Gem}s are conventionally cascaded to achieve a high gain with a negligible probability of sparking, but for our purpose the gain required is so low (well below 100) that a single \textsc{gem} suffices.
The material of one \textsc{gem} foil contributes $\sim 0.067$\%~X$_0$ to the material budget of the detector.

\subsection{Readout Circuit}
\begin{figure}[b]
\includegraphics[width=\columnwidth]{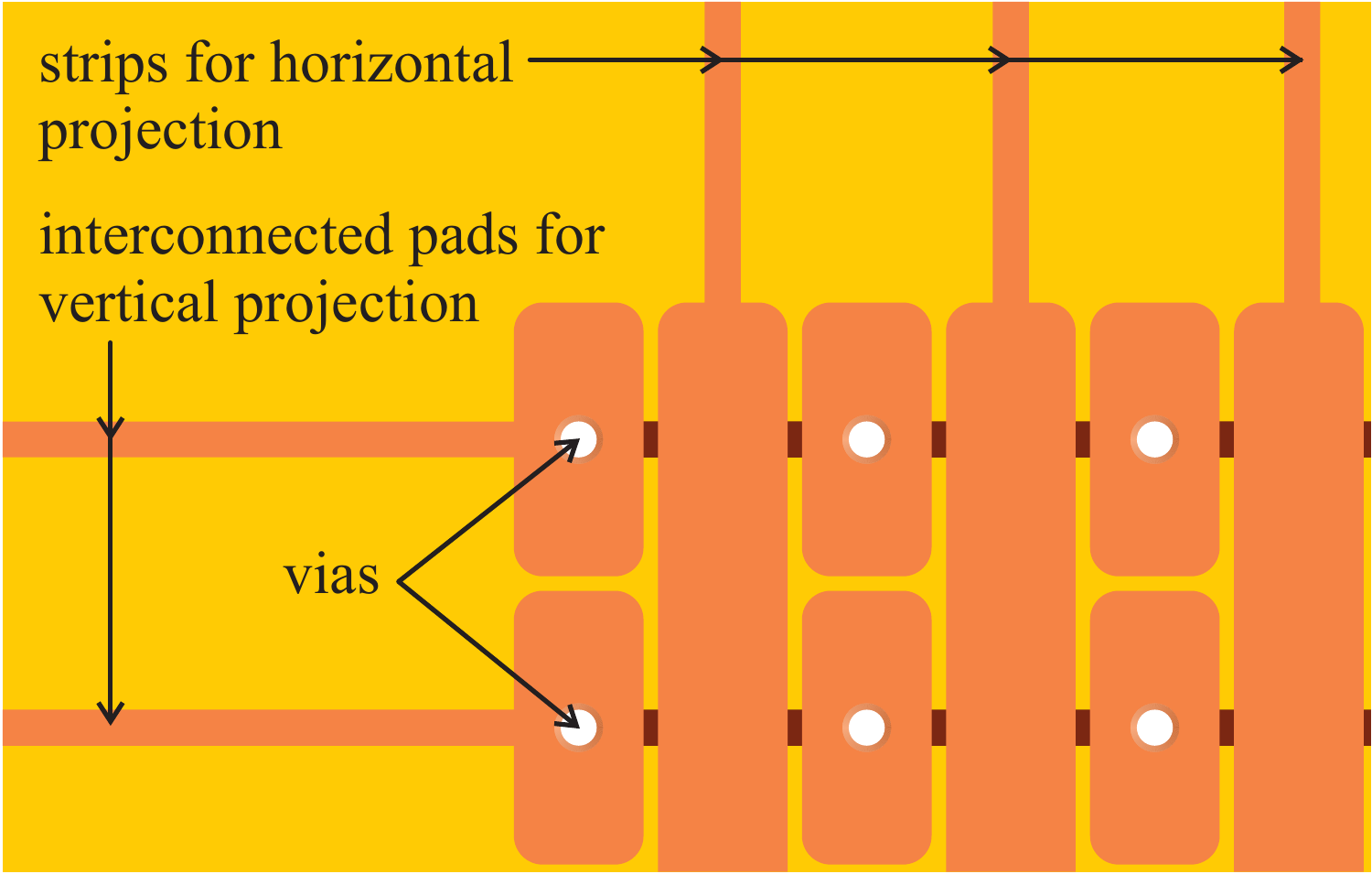}
\caption{Layout of a bidirectional readout circuit.}
\label{readout}
\end{figure}
The multiwire proportional chambers (\textsc{mwpc}s) used until now for measuring profiles read out horizontal and vertical profiles in separate chambers.
The chamber upstream causes so much multiple scattering to the beam that the profile measured by the downstream chamber is strongly degraded.
To avoid this situation with our \textsc{gem} chambers, we designed a readout circuit to read both projections in the same plane, see Fig~\ref{readout}.

It is a conventional double layer printed circuit board, with the readout elements on the top layer and signal routing traces on the bottom layer.
One projection of the profile is read out by strips, the other by rows of pads that are interconnected by traces on the bottom layer.
The collected charge is shared evenly between horizontal and vertical readout elements, and the channel density is the same horizontally and vertically.
This design can be im\-ple\-men\-ted on any base material; we chose to use polyimide (100 \textmu m thickness) in order to minimize the material budget.
The readout circuit is nevertheless the heaviest component of the detector, contributing $\sim 0.3\% \textrm{X}_0$.
Many techniques exist to make the many vias in such a design gas-tight, but we deliberately left them open so that gas can flow through.

\subsection{Gas Distribution and Thin Cathodes}
The distribution of gas trough the chamber is done by grooves milled in the fiberglass frames.
The gas flow is routed such that gas enters the chamber below the readout board, flows through the vias in the readout board and through the \textsc{gem} holes, and exits from the drift region.
This is indicated in Fig.~\ref{detector}.

The cathode is a crucial element when absorption and multiple scattering of the beam are of concern.
In our design the cathode is also the gas enclosure, and it is stretched tight ($\sim 11$ MPa) in order to avoid any deformation by the slight overpressure in the chamber.
It is made of the same base material \textsc{gem}s are made of: copperclad polyimide.
The copper is etched away in the active area of the detector, leaving just a thin ($\sim100$ nm) layer of chromium which is there to act as a tie coat for a better adhesion of the copper layer to the polyimide substrate.
The material traversed by the beam to enter the active volume of the detector thus amounts to 0.018\%~X$_0$.
On the other end of the chamber, the gas enclosure is made of a 25 \textmu m polyimide foil, adding 0.009\%~X$_0$.

\subsection{Electronics}
\begin{figure}
\includegraphics[width=\columnwidth]{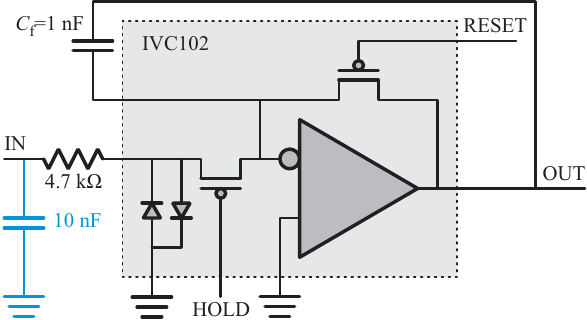}
\caption{Switched integrator circuit based on the IVC102.}
\label{integrator}
\end{figure}

The electronics used to read out the detector comes from the design used in other experimental areas at \textsc{cern}, but with some necessary modifications.
It is based on a conventional switched integrator circuit built around the IVC102 unit from Texas Instruments, see Fig.~\ref{integrator}.
After integrating during one second, amplitudes from 64 of such integrators (32 for each projection) are multiplexed and converted by an \textsc{adc}.
More details about the acquisition system are given in~\cite{Jens}.

The integrator has a large feedback capacitor (1 nF) to have a low sensitivity to noise and cross-talk.
This low sensitivity allows the integrators to be located outside the pendulum, about two meters of cable away from the detector.
Another consequence of the low sensitivity is that a lot of charge needs to be collected to reach the full scale of the \textsc{adc} that reads out the integrator ($V_\textrm{fs}=\pm 10$~V): $V_\textrm{fs}\cdot C_\textrm{f} = 10\cdot10^{-9} = 10$~nC per channel.
This charge is collected during a spill of a few hundred ns duration, giving rise to input currents of up to 100 mA.
The protection diodes indicated in Fig.~\ref{integrator} start clamping from an input current of 0.2 mA, because of the voltage drop over the series resistance of the \textsc{fet} switch marked \textsc{hold} ($\sim1.5$ k$\Omega$).
To limit the input current and avoid the non-linear behavior caused by the clamping diodes we added a capacitor to the input of each channel, indicated in blue in Fig.~\ref{integrator}.
This capacitor acts as a low-pass filter, collecting charge during a spill and then dissipating it slowly through the resistive elements into the integrator.
The effect of this modification can be seen in Fig.~\ref{profiles}, where profiles made with standard (left) and modified (right) electronics are shown for a range of high voltage settings.
At low amplitude there is no appreciable difference, but at high amplitude the modified integrators collect more than twice as much charge.
This is an easy way to use rather slow integrating electronics with beams with a fast spill structure.

\section{Results \& Discussion}
\begin{figure}[t]
\includegraphics[width=\columnwidth]{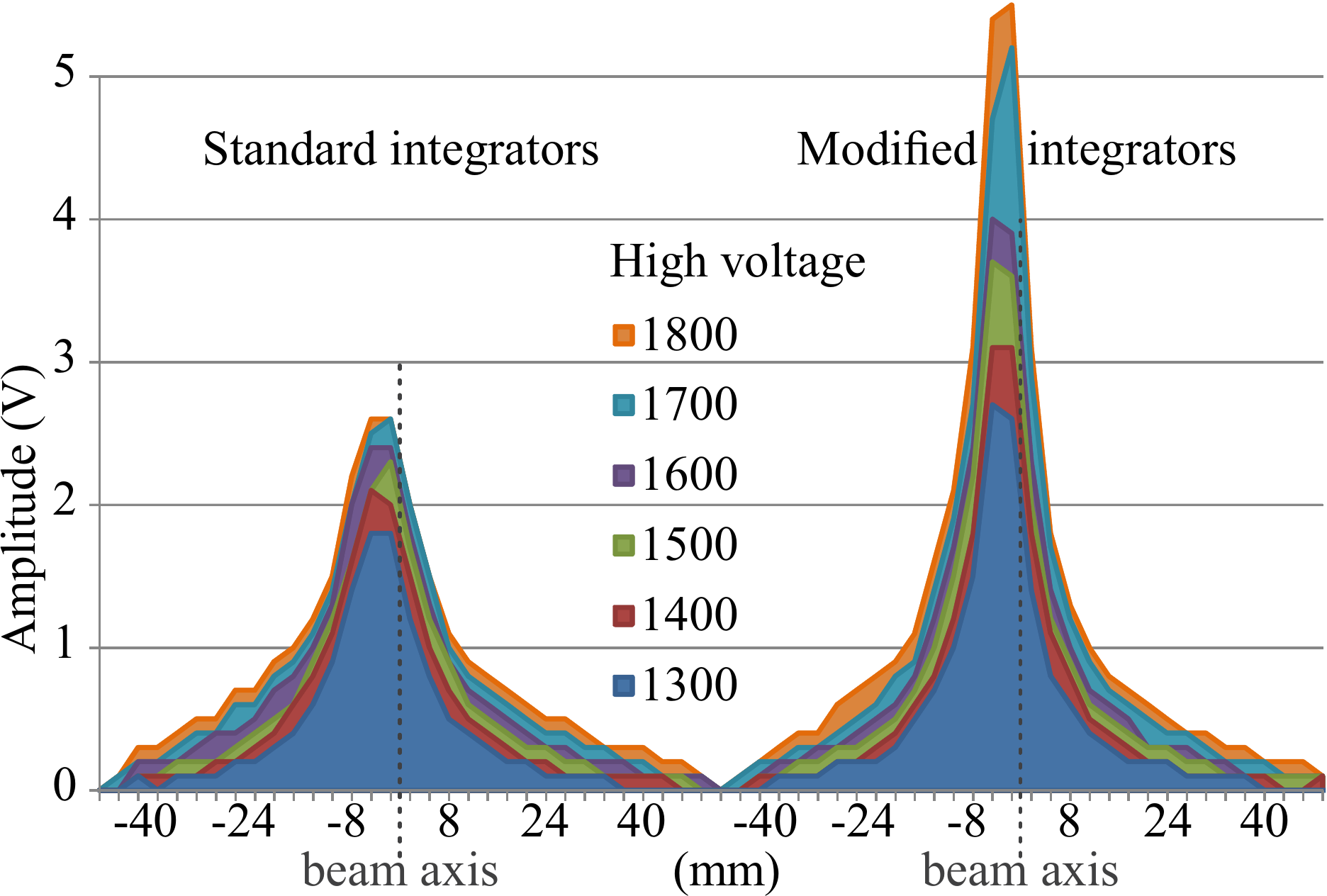}
\caption{Profiles made with standard (left) and modified (right) integrating electronics. The profiles are taken using a range of high voltage settings. The voltage indicated in the legend is the voltage of the cathode. The voltage over the \textsc{gem} is always $0.22\times$ the cathode voltage.}
\label{profiles}
\end{figure}

At the time of writing the detectors described above are installed in 6 locations in the \textsc{asacusa} and \textsc{aegis} beam lines, and replacement of the remaining \textsc{mwpc} monitors in all beam lines is foreseen for the end of 2011.
All these detectors give reliable, useful profiles for operators to steer the beam.
Figure~\ref{profiles} shows what these profiles look like, and how modifications of the electronics can improve the quality of the profiles.
Nevertheless, there are still aspects of performance that need improvement.

Lab tests done with the detector of Fig.~\ref{profiles} before it was installed showed a much steeper dependence of gain on high voltage than what can be seen in the figure.
It seems like the integrators are still not working entirely linearly, especially at higher amplitudes.
A possible cure would be to increase the sensitivity of the integrators (i.e. reduce the value of $C_\mathrm{f}$ in Fig.~\ref{integrator}), as far as noise allows.
This would further reduce the instantaneous input current into the integrator circuit, and also the current the \textsc{gem} needs to drive.

For the highest profiles of Fig.~\ref{profiles} we estimate that the charge on the capacitance of a \textsc{gem} ($\sim4.7$ nF) is reduced by about 70 nC during a spill.
This gives rise to a drop in \textsc{gem} voltage of $Q/C_\mathrm{GEM}=70/4.7\approx 15$ V, which according to our gain calibration data corresponds to a factor of 2.2 reduction in gain.
This contributes to the reduced linearity at higher amplitudes we observe, although it cannot fully explain it.

Recent tests with a beam collimated to 3 mm gave profiles as wide as 10 mm \textsc{fwhm}.
This widening is probably caused by multiple scattering of the low energy beam in the vacuum window, 4 cm upstream the detector.
To improve the spatial resolution the detector must therefore be installed as close as possible to the window.
Ultimately, the vacuum windows may need to be replaced by lighter or thinner foils.

\section{Conclusion \& Outlook}
We developed new transverse beam profile monitors for the \textsc{cern ad} beam lines, and report on the first results with these detectors.
The monitors are single \textsc{gem} detectors with a bidirectional readout structure, and the chamber is designed to minimize material budget.
They dramatically improve the profile measurements compared to the \textsc{mwpc}s currently used, even if there is room for improvement.

These detectors are very easy to operate and maintain, and based on inexpensive components.
This makes detectors of a similar design likely candidates for replacement of \textsc{mwpc}s currently in use in many higher energy beam lines at \textsc{cern}.


\vfill
\begin{thebibliography}{9}

\bibitem{AD}
Lajos Bojt\'ar, on behalf of the \textsc{ad} team, ``Antiproton Decelerator Status Report'',  \textsc{Cool'09},  Lanzhou , P.R.China.

\bibitem{firstGEM}
F. Sauli, ``\textsc{Gem}: A new concept for electron amplification in gas detectors'', Nucl. Inst. \&  Meth. A 386, 2--3, 1997.

\bibitem{Jens}
J. Spanggaard et al., ``Gas Electron Multipliers for low Ener\-gy Beams'', \textsc{Ipac'10}, Kyoto, Japan.

\end{thebibliography}
\end{document}